# Quantum Stochastic Stability

Javad Sharifi , Hamidreza Momeni

Automatic Control Laboratory, Electrical Engineering Department, Tarbiat Modares University, Tehran, Iran

jv.sharifi@gmail.com , momeni_h@modares.ac.ir

**Abstract**—the stability criterion is constructed for open quantum systems which govern by quantum stochastic differential equations (QSDE) both for quantum observable flow and the stochastic density operator. We derive stability criteria (local, asymptotic and exponential stability) for those QSDE.

**Keywords**: quantum stochastic differential equations, quantum stochastic stability, quantum flow, stochastic density operator

**1. Introduction**

Quantum stochastic differential equation (QSDE) describes the evolution of interacted quantum systems with fundamental quantum noises. Quantum stochastic calculus and quantum Itô formula which goes back to 1984 and was basically developed by Hudson and Parthasarathy [1]. They extract some stochastic evolution for open quantum systems and their model is a good model in quantum optics [2].

The deterministic stability theory in first was introduced by Lyapunov, and studies the behavior of physical and dynamical systems and is an essential part of control systems. Classical stochastic analogue of Lyapunov stability is well developed by Kushner [3] and Has'minski [4] and Mao [5] which the stochastic processes derive with classical fundamental noise processes, i.e. Wiener and Poisson process.

The equation of quantum trajectory (quantum filtering), which was developed by Belavkin, is a classical Itô process. The stability analysis of quantum trajectories, in our knowledge, at first appears in [6] which especially was solved in detail for two important quantum systems, the first one is ensemble of atoms angular momentum and the second is the two spin entanglement that are very important in quantum computing.

In [7] the stability criterion is developed for Markovian quantum filter in the sense that any randomizations of initial density operator get a true estimate of system state in infinite time horizon. Likewise, a local stability criterion is extracted for quantum Markov process using quantum Dynkin's formula in [8].

Here we derive some stability theorems for QSDE which evolves with fundamental quantum noises, i.e. annihilation, creation and gauge process, both for quantum stochastic flow and stochastic density matrix. To do this, we briefly review the quantum probability and quantum stochastic process. After that, we define some quantum stochastic stability conditions and prove several theorems for them.

## 2. Quantum Stochastic Process

### 3.1. Quantum Probability and Statistics

Let $\mathfrak{H}$ be some complex Hilbert space with inner product $\langle,\rangle$ and norm $\|\psi\|^2 = \langle \psi, \psi \rangle$. Let $\mathbf{B}(\mathfrak{H})$ be a C*-algebra[1] of bounded operators over $\mathfrak{H}$ and $\mathcal{A} \subset \mathbf{B}(\mathfrak{H})$ be a subalgebra. Let $\mathbb{P}: \mathcal{A} \to \mathbb{C}$ be a positive linear functional with $\mathbb{P}(\mathbf{I}) = 1$ for identity operator[2] $\mathbf{I}$, then $(\mathcal{A}, \mathbb{P})$ is a quantum probability space. The norm of an operator is $\|X\| \triangleq \sup\{\|X\psi\| : \psi \in \mathfrak{H}, \|\psi\| = 1\}$. $\|X\| \geq 0$ means that the operator is positive which we also show as $X \geq 0$ and $X > 0$ is strictly positive operator. We have the relation $|\mathbb{P}(X)| \leq \|X\|$ for quantum probability of an operator and its norm and for self-adjoint operator $X^* = X: \mathbb{P}(X) \geq 0$ [9].

Quantum state enables the computation of statistical properties of quantum operators. A state in a separable Hilbert space $\mathfrak{H}$, denotes by a trace one positive density operator $\rho = |\psi\rangle\langle\psi|$ where $\psi \in \mathfrak{H}$. The (vacuum) expectation of quantum operators $X \in \mathcal{A}$ via quantum state is: $\langle X \rangle_\rho = \mathbb{E}_\rho(X) \triangleq \mathrm{Tr}(\rho X)$. For further reading about quantum probability see [9, 13, 15].

### 3.2. Quantum Noise on Fock Space

Hudson and Parthasarathy obtained a Fock-space representation of Wiener and Poisson Processes which play an important role in quantum stochastic calculus. Let $\mathfrak{H}_F = L^2(\mathbb{R}_+)$ be a Hilbert space, then the symmetric Fock space $\mathcal{F}(\mathfrak{H}_F)$ is defined as $\mathcal{F}(\mathfrak{H}_F) = \mathbb{C} \oplus \bigoplus_{n=1}^{\infty} \mathfrak{H}_F^{\odot n}$, where $\odot$ is the symmetric tensor product. In fact, the symmetric Fock space $\mathcal{F}(\mathfrak{H}_F)$ is the Hilbert space completion of the linear span of the exponential vectors $e(\mathfrak{f}) \triangleq 1 \oplus \bigoplus_{n=1}^{\infty} \frac{1}{\sqrt{n!}} \mathfrak{f}^{\odot n}$ under the inner product $\langle e(\mathfrak{f}), e(\mathfrak{g}) \rangle = \exp(\langle \mathfrak{f}, \mathfrak{g} \rangle)$ where $\mathfrak{f}, \mathfrak{g} \in L^2(\Omega, \mathbb{C})$, $\Omega \subset \mathbb{R}_+$ and $\langle \mathfrak{f}, \mathfrak{g} \rangle \triangleq \int_0^{+\infty} \mathfrak{f}^*(t)\mathfrak{g}(t)dt$. Likewise, $e(0)$ is the vacuum vector. Quantum noises are *annihilation* ($A_t^\dagger = A_{\chi_{[0,t]}}^\dagger$), *creation* ($A_t = A_{\chi_{[0,t]}}$) and *gauge* ($\Lambda_t$) operators by their action on exponential vector as [1]:

$$A_{(t)}e(\mathfrak{f}) = \left(\int_0^t \mathfrak{f}(s)ds\right)e(\mathfrak{f}) \ , \ A_{(t)}^\dagger e(\mathfrak{f}) = \frac{d}{d\delta}e(\mathfrak{f} + \delta\chi_{[0,t)})\bigg|_{\delta=0} \ , \ \Lambda_{(t)}e(\mathfrak{f}) = \frac{d}{d\delta}e(\exp(\delta\chi_{[0,t)})\mathfrak{f})\bigg|_{\delta=0} \tag{1}$$

### 3.3. QSDE and Stochastic Calculus

Quantum Itô rule due to Hudson-Parthasarathy play an important rule in quantum stochastic calculus and stochastic evolution. We mention the fundamental theorems and equations from [1, 10] that is essential for derivation of quantum stochastic stability:

---

[1] A C*-algebra $\mathbf{B}(\mathfrak{H})$ is a complex normed algebra which is complete with respect to its norm and for two operators $P, Q \in \mathbf{B}(\mathfrak{H})$ satisfies the relations: $\|PQ\| \leq \|P\|\|Q\|$ and $\|P^*P\| = \|P\|^2$.

[2] $\mathbf{I}|\varsigma\rangle = |\varsigma\rangle$, $\langle\varsigma|\mathbf{I} = \langle\varsigma|$ ; $\forall \varsigma \in \mathfrak{H}$

**Theorem 2.1** (quantum Itô rule): Let $dX_t, dY_t$ be the quantum stochastic processes as $dX_t = C_t dA_t + D_t dA_t^\dagger + E_t d\Lambda_t + F_t dt$ and $dY_t = G_t dA_t + H_t dA_t^\dagger + K_t d\Lambda_t + L_t dt$ then for some stochastically integrable processes $C, D, E, F, G, H, K, L$ the stochastic process $X_t Y_t$ satisfies the multiplication rule: $d(X_t Y_t) = d(X_t) Y_t + X_t d(Y_t) + d(X_t) d(Y_t)$, which for calculations we must use the quantum Itô table: $d\Lambda_t d\Lambda_t = d\Lambda_t, d\Lambda_t dA_t^\dagger = dA_t^\dagger, dA_t d\Lambda_t = dA_t, dA_t dA_t^\dagger = dt$ with all the other products being equal to zero.

*Quantum Stochastic Evolution*: The Hudson-Parthasarathy QSDE for unitary operator $U_t \in \mathcal{S} \otimes \mathcal{W}$, i.e. $U_t U_s = U_{t+s}, U_{-s} = U_s^*, U_0 = U_s U_s^* = U_s^* U_s = I$, is:

$$dU_t = \left\{ \left(-iH - \tfrac{1}{2} L^* L\right) dt + L dA_t^\dagger - L^* S dA_t + (S - I) d\Lambda_t \right\} U_t, \quad U_0 = \mathbf{I} \qquad (2)$$

where $S, H, L \in \mathcal{S} \otimes \mathcal{W}$ are bounded operators. Hamiltonian $H$ is self-adjoint and $S$ is scattering operator which is unitary, $SS^* = S^* S = I$. This equation has many applications in quantum optics and it is the start point for derivation of quantum filtering equation [11-15].

## 4. Quantum Stochastic Stability

The stability of quantum stochastic process which we develop here, studies the behavior of quantum observable flow and the stochastic evolution of density operator. We prove three of main theorems for stability of quantum flow. The stability theorem of stochastic density operator is similar to quantum flow and the proof is identical, hence we omit, but quantum flow and density operator have two different QSDE.

### 4.1. Stability of Quantum Flow

*Quantum flow* for operator process $X_t = U_t^* X U_t \triangleq f_t(X)$, $X = X_0$, using quantum Itô rule, is:

$$dX_t = df_t(X) = f_t(\mathfrak{L}_{L,H}(X)) dt + f_t([L^*, X] S) dA_t + f_t(S^* [X, L]) dA_t^\dagger + f_t([S^* X, S]) d\Lambda_t \qquad (3)$$

where $\mathfrak{L}_{L,H}(X) \triangleq i[H, X] + L^* X L - \tfrac{1}{2}\{L^* L, X\}$ is the generator of quantum flow [1,10] and $[A, B] = AB - BA$, $\{A, B\} = AB + BA$. In this section, we derive the stability condition for this type of QSDE.

**Definition1:** a *quantum stop time* is a self-adjoint operator $\tau = \int_{\mathbb{R}^+ \cup \{\infty\}} \lambda \, d\mathbf{S}(\lambda)$ with spectral resolution $\mathbf{S}(\lambda) : \lambda \in \mathbb{R}^+$, and commutes with quantum flow: $[\tau([0,t]), f_s(X)] = 0 : s \geq t$ which means that observation of stop process at $\tau([0,t])$ does not interference with future quantum flow or physically non-demolish with respect to quantum flow [10].

For commutative quantum stochastic process, the quantum stop time is a measurable classical stop time. The projection $\tau([0,t])$ means the stopping the quantum stochastic process (flow) occurred at or before time $t$. For example, the first exit time of quantum flow from a set defines as a quantum stop time which we use in the sequel.

**Definition2:** suppose $X_e$ is equilibrium for quantum flow (3), i.e. $f_t(\mathfrak{L}_{L,H}(X_e)) = f_t([L^*, X_e] S) = f_t(S^* [X_e, L]) = f_t([S^* X_e, S]) = 0$; $f_t(X_e) = X_e$, Then the quantum operator $X_e$ is:

1- *Locally quantum Stable* if $\forall \alpha > 0$ and $\beta \in (0,1)$, there is $\delta > 0$ such that for a positive operator function $\mathbf{V}(X_t)$ we have $\mathbb{P}\left(\underset{t \geq 0}{\text{Sup}}\, \mathbf{V}(X_{t \wedge \tau}) > \alpha \mathbf{I}\right) \leq \beta$ for some initial $\delta$-neighborhood of $X_e$, i.e. $\mathbf{V}(X) < \delta \mathbf{I}$.

2- *Asymptotically quantum stable* if it is stable in quantum probability and also $\underset{X \to X_e}{\lim} \mathbb{P}\left(\underset{t \to \infty}{\lim}\, \mathbf{V}(X_t) = 0\mathbf{I}\right) = 1$

3- *Exponentially quantum stable* if it is stable in quantum probability and for all $T < \infty$: $\mathbb{P}\left(\underset{t \geq 0}{\text{Sup}}\, \mathbf{V}(X_t) > \alpha \mathbf{I}\right) \leq \beta \exp(-at)$, where $\infty > \alpha, \beta, a > 0$

The local stability means that with initial condition $X$ in $\delta$-neighborhood of equilibrium observable $X_e$, the observable evolution $X_t$ remains in $\alpha$-neighborhood with quantum probability no less than $\beta \in (0,1)$. The asymptotic stability means, the observable reach to equilibrium observable. The exponential stability indicates on the rate of convergence.

First of all, let us prove a useful lemma:

**Lemma 3.1**: let $X$ be bounded quantum operator, i.e. $\exists 0 < \alpha < \infty : \|X\| \leq \alpha$, and $\Theta$ be a bounded finite-dimensional operator, then for every natural numbers $n, m$ the operator $X^n \Theta X^m$ is bounded, i.e. $\exists \gamma > 0 : \|X^n \Theta X^m\| \leq \gamma$.

**Proof:** note that we define the norm of operator as: $\|X\| \triangleq \text{Sup}\{\|X\varsigma\| : \varsigma \in \mathfrak{H}, \|\varsigma\| = 1\}$ then the by general definition of bounded operators is $\exists 0 < \alpha < \infty : \|X\varsigma\| \leq \alpha \|\varsigma\|$ which by norm definition, we get $\|X\| \leq \alpha$ and $\|\Theta\| \leq \beta$. For bounded operators $X, Y$ there is inequality [16]: $\|XY\| \leq \|X\|\|Y\|$, which by this relation we get the result: $\|X^n \Theta X^m\| \leq \|X\|^n \|\Theta\| \|X\|^m \leq \alpha^{n+m} \beta \triangleq \gamma$. $\square$

Now, consider a bounded quantum operator value function $\mathbf{V}(X_t)$. Without loss of generality, assume it is as arbitrary finite dimension polynomial span of operator $X_t$ as $\mathbf{V}(X_t) = \sum_{n,m} X_t^n \Theta_{n,m} X_t^m \equiv X_t^n \Theta_{n,m} X_t^m$ where $m, n \in \mathbb{N} + \{0\}$ and $\Theta_{n,m}$ are some bounded operator (matrices) with appropriate dimensions, then by lemma1 $\mathbf{V}(X_t)$ is bounded. Here, we want calculate the differential $d\mathbf{V}(X_t) = d\mathbf{V}(U_t^* X U_t)$. For this purpose, note that since $U_t$ is unitary operator, then we obtain the following facts:

$$X_t Y_t = f_t(X) f_t(Y) = U_t^* X U_t U_t^* Y U_t = f_t(XY), \quad X_t + Y_t = U_t^*(X+Y) U_t = f_t(X+Y) \tag{4}$$

We have: $\mathbf{V}(X_t) = f_t(X^n) \Theta_{n,m} f_t(X^m)$. Then by quantum Itô rule and using equations (2) and (3), easily obtains:

$$d\mathbf{V}(X_t) = df_t(X^n) \Theta_{n,m} f_t(X^m) + f_t(X^n) \Theta_{n,m} df_t(X^m) + df_t(X^n) \Theta_{n,m} df_t(X^m) = \aleph_{(t)} dt + \aleph_{(A)} dA_t + \aleph_{(A^\dagger)} dA_t^\dagger + \aleph_{(\Lambda)} d\Lambda_t \tag{5}$$

Where the operators $\aleph_{(t)}, \aleph_{(A)}, \aleph_{(A^\dagger)}, \aleph_{(\Lambda)}$ for this function are:

$$\begin{aligned}
\aleph_{(t)} &= f_t\left(\mathcal{L}_{L,H}(X^n)\right)\Theta_{n,m}f_t(X^m) + f_t(X^n)\Theta_{n,m}f_t\left(\mathcal{L}_{L,H}(X^m)\right) + f_t\left(\left[L^*,X^n\right]S\right)\Theta_{n,m}f_t\left(S^*\left[X^m,L\right]\right) \\
\aleph_{(A)} &= f_t\left(\left[L^*,X^n\right]S\right)\Theta_{n,m}\left\{f_t(X^m) + f_t\left(\left[S^*X^m,S\right]\right)\right\} + f_t(X^n)\Theta_{n,m}f_t\left(\left[L^*,X^m\right]S\right) \\
\aleph_{(A^\dagger)} &= \left\{f_t(X^n) + f_t\left(\left[S^*X^n,S\right]\right)\right\}\Theta_{n,m}f_t\left(S^*\left[X^m,L\right]\right) + f_t\left(S^*\left[X^n,L\right]\right)\Theta_{n,m}f_t(X^m) \\
\aleph_{(\Lambda)} &= \left\{f_t(X^n) + f_t\left(\left[S^*X^n,S\right]\right)\right\}\Theta_{n,m}f_t\left(\left[S^*X^m,S\right]\right) + f_t\left(\left[S^*X^n,S\right]\right)\Theta_{n,m}f_t(X^m)
\end{aligned} \qquad (6)$$

We always for every other bounded positive operator value function $\mathbf{V}(X_t)$, the expression like last part of equation (5). By those definitions, we know ready to prove some quantum stochastic stability theorems:

**Theorem 3.1** (*Locally Quantum Stable*): consider the QSDE in (3) and define the set $\mathbb{Q}_\varepsilon \triangleq \{X_t : \mathbf{V}(X_t) \leq \varepsilon \mathbf{I}\}$; if there is positive operator function $\mathbf{V}(X_t)$ such that $\mathbf{V}(X_e) = 0$, $\mathbf{V}(X_t) > 0 : \forall X_t \in \mathbb{Q}_\varepsilon - \{X_t\}$ and $\aleph_{(t)} \leq 0 : \forall X_t \in \mathbb{Q}_\varepsilon$, then the quantum observable $X_e$ is stable in sense of quantum probability.

**Proof**: we wish to make $\mathbb{P}\left(\underset{t\geq 0}{\text{Sup}}\,\mathbf{V}(X_{\tau \wedge t}) \geq \varepsilon \mathbf{I}\right)$ arbitrarily small for sufficiently small initial distance $\mathbf{V}(X) < \delta \mathbf{I}$. For some non-demolish quantum stop-time $\tau \triangleq \inf\{t : X_t \notin \mathbb{Q}_\varepsilon\}$, by quantum Itô rule, (see equation (5)), we find that:

$$\mathbf{V}(X_{t\wedge\tau}) = \mathbf{V}(X) + \int_0^{t\wedge\tau}\aleph_{(s)}ds + \int_0^{t\wedge\tau}\aleph_{(A)}dA_s + \int_0^{t\wedge\tau}\aleph_{(A^\dagger)}dA_s^\dagger + \int_0^{t\wedge\tau}\aleph_{(\Lambda)}d\Lambda_s \qquad (7)$$

From the fact that the vacuum expectation of quantum stochastic integrals is zero, see [17], it obtains $\mathbb{E}_\rho\left(\mathbf{V}(X_{t\wedge\tau})\right) = \mathbb{E}_\rho\left(\mathbf{V}(X)\right) + \int_0^{t\wedge\tau}\mathbb{E}_\rho\left(\aleph_{(s)}\right)ds$. For $\aleph_{(t)} \leq 0$, the time integral is non-increasing. Without loss of generality suppose for $\Psi = \xi \otimes \zeta : \xi \in \mathfrak{H}_S, \zeta \in \mathcal{F}$ the quantum expectation be pure state expectation, Since expectation could be expressed as a convex combination of such vector inner products [18], then we find: $\mathbb{E}_\rho\left(\mathbf{V}(X_{t\wedge\tau})\right) \triangleq \langle\Psi, \mathbf{V}(X_{t\wedge\tau})\Psi\rangle \leq \langle\Psi, \mathbf{V}(X)\Psi\rangle = \mathbb{E}_\rho\left(\mathbf{V}(X)\right)$, By this inequality together with Chebyshev's inequality, we obtain: $\mathbb{P}\left(\underset{t\geq 0}{\text{Sup}}\,\mathbf{V}(X_{t\wedge\tau}) > \alpha\mathbf{I}\right) \leq \alpha^{-1}\mathbb{E}_\rho\left(\mathbf{V}(X_{t\wedge\tau})\right) \leq \alpha^{-1}\mathbb{E}_\rho\left(\mathbf{V}(X)\right)$. By this last inequality and the relation $\mathbb{E}_\rho\left(\mathbf{V}(X)\right) < \delta\mathbb{E}_\rho(\mathbf{I}) = \delta\langle\Psi,\mathbf{I}\Psi\rangle = \delta\langle\Psi,\Psi\rangle = \delta$, then conclude $\mathbb{P}\left(\underset{t\geq 0}{\text{Sup}}\,\mathbf{V}(X_{t\wedge\tau}) > \alpha\mathbf{I}\right) < \alpha^{-1}\delta \triangleq \beta$. Note that $\beta$ could become arbitrarily small by choosing initial condition $X$ sufficiently close to $X_e$. □

**Theorem 3.2** (*Asymptotically Quantum Stable*): consider quantum flow (3); if there is a bounded positive operator value function $\mathbf{V}(X_t)$ such that $\mathbf{V}(X_e) = 0$, $\mathbf{V}(X_t) > 0$, $\aleph_{(t)} < 0 : \forall X_t \in \mathbb{Q}_\varepsilon - \{X_e\}$, then the quantum observable equilibrium $X_e$ is asymptotically open quantum stable.

**Proof**: we want to prove that with sufficiently close initial condition to desired operator, we have convergence in quantum probability to equilibrium, i.e. $\underset{X\to X_e}{\text{Lim}}\mathbb{P}\left(\underset{t\to\infty}{\text{Lim}}\mathbf{V}(X_t) > 0\mathbf{I}\right) = 0$. Define the level sets: $\mathbb{Q}_{\varepsilon_i} \triangleq \{X_t : \mathbf{V}(X_t) \leq \varepsilon_i\mathbf{I}\} : \varepsilon_i \in \mathbb{R}^+$, $\mathbb{Q}_{\varepsilon_{i+1}} \subset \mathbb{Q}_{\varepsilon_i}$, $\varepsilon_i > \varepsilon_{i+1}$, $i = 1,..,n$ and the quantum stop times: $\tau_i \triangleq \inf\{t : X_t \notin \mathbb{Q}_{\varepsilon_i}\} : \tau_n > \tau_{n-1} > \cdots > \tau_1 \geq 0$, then using the quantum Itô rule and taking vacuum expectation:

$$\mathbb{E}_\rho\left(\mathbf{V}(X_{t\wedge\tau_{i+1}})\right) = \mathbb{E}_\rho\left(\mathbf{V}(X_{t\wedge\tau_i})\right) + \int_{t\wedge\tau_i}^{t\wedge\tau_{i+1}} \mathbb{E}_\rho\left(\aleph_{(t)}\right)dt : \tau_{i+1} > \tau_i, [\tau_i,\tau_{i+1}] = 0 \qquad (8)$$

By assumption $\mathbb{E}_\rho\left(\aleph_{(t)}\right) < 0$, for two stop times $\tau_i < \tau_{i+1}: \mathbb{E}_\rho\left(\mathbf{V}(X_{t\wedge\tau_{i+1}})\right) < \mathbb{E}_\rho\left(\mathbf{V}(X_{t\wedge\tau_i})\right)$ and similarly for sequences of stop times $0 < \tau_1 < \tau_2 < \cdots < \tau_n$, we find that: $\mathbb{E}_\rho\left(\mathbf{V}(X_{t\wedge\tau_n})\right) < \cdots < \mathbb{E}_\rho\left(\mathbf{V}(X_{t\wedge\tau_2})\right) < \mathbb{E}_\rho\left(\mathbf{V}(X_{t\wedge\tau_1})\right) < \mathbb{E}_\rho\left(\mathbf{V}(X)\right)$. We now, obtain the condition such that the quantum observable flows from one level set to another in finite time. For this purpose suppose that $\aleph_{(t)} \leq -b\mathbf{I} < 0$, $b \in \mathbb{R}^+$, then by equation (8), we obtain:

$$\mathbb{E}_\rho\left(\mathbf{V}(X_{t\wedge\tau_{i+1}})\right) - \mathbb{E}_\rho\left(\mathbf{V}(X_{t\wedge\tau_i})\right) = \int_{t\wedge\tau_i}^{t\wedge\tau_{i+1}} -\mathbb{E}_\rho\left(\aleph_{(t)}\right)dt > b\mathbb{E}_\rho\left(\int_{t\wedge\tau_i}^{t\wedge\tau_{i+1}} \mathbf{I}dt\right) = b\mathbb{E}_\rho\left(t\wedge\tau_{i+1} - t\wedge\tau_i\right) \qquad (9)$$

then $\mathbb{E}_\rho\left(t\wedge\tau_{i+1} - t\wedge\tau_i\right) < b^{-1}\left(\mathbb{E}_\rho\left(\mathbf{V}(X_{t\wedge\tau_{i+1}})\right) - \mathbb{E}_\rho\left(\mathbf{V}(X_{t\wedge\tau_i})\right)\right) < b^{-1}\mathbb{E}_\rho\left(\mathbf{V}(X_{t\wedge\tau_{i+1}})\right)$ and with bounded $\mathbf{V}(X_{t\wedge\tau_{i+1}})$, the quantum observable flows from $\mathbb{Q}_{\varepsilon_i}$ to $\mathbb{Q}_{\varepsilon_{i+1}}$ in a finite average time. Likewise, as $t \to \infty$, we get for some sample path $\tau_n \to \tau = \infty: \mathbb{E}_\rho\left(\mathbf{V}(X_\infty)\right) < \mathbb{E}_\rho\left(\mathbf{V}(X_{\tau_1})\right)$. Then with similar argument in previous proof, implies $\mathbb{P}\left(\lim_{t\to\infty}\mathbf{V}(X_t) \geq \alpha\mathbf{I}\right) < \alpha^{-1}\mathbb{E}_\rho\left(\mathbf{V}(X)\right)$. For those sample path: $\int_0^\infty -\mathbb{E}_\rho\left(\aleph_{(t)}\right)dt = \mathbb{E}_\rho\left(\mathbf{V}(X)\right) - \lim_{s\to\infty}\mathbb{E}_\rho\left(\mathbf{V}(X_s)\right)$ which with $\mathbf{V}(X_t)$ bounded and by assumption $-\mathbb{E}_\rho\left(\aleph_{(t)}\right) > 0$, the statement $\int_0^\infty -\mathbb{E}_\rho\left(\aleph_{(t)}\right)dt < \infty$ satisfies at least for $\lim_{t\to\infty}\mathbb{E}_\rho\left(\aleph_{(t)}\right) = 0$. □

**Theorem 3.3** (*Exponentially Quantum Stable*): if there is a bounded positive operator value function $\mathbf{V}(X_t)$ such that $\mathbf{V}(X_e) = 0$, $\mathbf{V}(X_t) > 0$, $\aleph_{(t)} + a\mathbf{V}(X_t) < 0 : \forall X_t \in \mathbb{Q}_\varepsilon - \{X_e\}$, $a > 0$, then the observable equilibrium $X_e$ is exponentially stable for quantum flow (3).

**Proof**: from previous proof, we know $\mathbb{E}_\rho\left(\mathbf{V}(X_{t\wedge\tau})\right) = \mathbb{E}_\rho\left(\mathbf{V}(X)\right) + \int_0^{t\wedge\tau} \mathbb{E}_\rho\left(\aleph_{(s)}\right)ds < \mathbb{E}_\rho\left(\mathbf{V}(X)\right) + \int_0^{t\wedge\tau} -a\mathbb{E}_\rho\left(\mathbf{V}(X_s)\right)ds$, which implies, by Gronwall lemma, that: $\mathbb{E}_\rho\left(\mathbf{V}(X_{t\wedge\tau})\right) < \mathbb{E}_\rho\left(\mathbf{V}(X)\right)\exp(-a(t\wedge\tau))$ and finally the result follows as:

$$\mathbb{P}\left(\sup_{t\geq 0}\mathbf{V}(X_{t\wedge\tau}) > \alpha\mathbf{I}\right) \leq \alpha^{-1}\mathbb{E}_\rho\left(\mathbf{V}(X_{t\wedge\tau})\right) \leq \alpha^{-1}\mathbb{E}_\rho\left(\mathbf{V}(X)\right)\exp(-a(t\wedge\tau)) = \beta\exp(-a(t\wedge\tau)). \Box$$

**Remark**: The exponential quantum stability estimates the probability rate that the quantum observable flows to equilibrium.

### 4.2. Stability of Quantum State

As it known, the quantum state in quantum mechanics is denoted by a self-adjoint density operator $\rho_t$. Herein, we investigate the stability of quantum state. The unitary evolution of quantum state is [2, 16] $\rho_t = U_t\rho U_t^*$, $\rho = \rho_0$ which by using the quantum Itô rule: $d\rho_t = dU_t\rho U_t^* + U_t\rho dU_t^* + dU_t\rho dU_t^*$, we find the following QSDE for stochastic density operator:

$$d\rho_t = \mathscr{L}(\rho_t)dt + \left(\left[\rho_t, L^*S\right]S^*\right)dA_t + \left(S\left[S^*L, \rho_t\right]\right)dA_t^\dagger + \left[S\rho_t, S^*\right]d\Lambda_t \tag{10}$$

where $\mathscr{L}(\rho_t) \triangleq -i[H, \rho_t] + L^*S\rho_t S^*L - \frac{1}{2}\{L^*L, \rho_t\}$. Let denote an equilibrium of quantum state by $\rho_e$ which obtain by setting all the differential coefficient in (10) being equal to zero or equivalently: $\rho_e = U_t \rho_e U_t^*$. Without loss of generality, let us set: $\mathbf{V}(\rho_t) = \sum_{n,m} \rho_t^n \Theta_{n,m} \rho_t^m \equiv \rho_t^n \Theta_{n,m} \rho_t^m : n, m \in \mathbb{N} + \{0\}$. Since $\rho_t^n = U_t \rho^n U_t^*$ then by quantum Itô rule, we obtain:

$$d\rho_t^n = \mathscr{L}(\rho_t^n)dt + \left(\left[\rho_t^n, L^*S\right]S^*\right)dA_t + \left(S\left[S^*L, \rho_t^n\right]\right)dA_t^\dagger + \left[S\rho_t^n, S^*\right]d\Lambda_t \tag{11}$$

and likewise:

$$d\left(\rho_t^n \Theta_{n,m} \rho_t^m\right) = d\left(\rho_t^n\right)\Theta_{n,m}\rho_t^m + \rho_t^n \Theta_{n,m} d\left(\rho_t^m\right) + d\left(\rho_t^n\right)\Theta_{n,m} d\left(\rho_t^m\right) \tag{12}$$

by equations (11) and (12) we get:

$$\begin{aligned}d\mathbf{V}(\rho_t) = \aleph_{(\rho,t)}dt &+ \left(\left[\rho_t^n, L^*S\right]S^*\Theta_{nm}\rho_t^m + \rho_t^n\Theta_{nm}\left[\rho_t^m, L^*S\right]S^* + \left[\rho_t^n, L^*S\right]S^*\Theta_{nm}\left[S\rho_t^m, S^*\right]\right)dA_t \\ &+ \left(S\left[S^*L, \rho_t^n\right]\Theta_{nm}\rho_t^m + \rho_t^n\Theta_{nm}S\left[S^*L, \rho_t^m\right] + \left[S\rho_t^n, S^*\right]\Theta_{nm}S\left[S^*L, \rho_t^m\right]\right)dA_t^\dagger \\ &+ \left(\rho_t^n\Theta_{nm}\left[S\rho_t^m, S^*\right] + \left[S\rho_t^n, S^*\right]\Theta_{nm}\rho_t^m + \left[S\rho_t^n, S^*\right]\Theta_{nm}\left[S\rho_t^m, S^*\right]\right)d\Lambda_t\end{aligned} \tag{13}$$

where $\aleph_{(\rho,t)} = \mathscr{L}(\rho_t^n)\Theta_{nm}\rho_t^m + \rho_t^n\Theta_{nm}\mathscr{L}(\rho_t^m) + \left[\rho_t^n, L^*S\right]S^*\Theta_{nm}S\left[S^*L, \rho_t^m\right]$ and by using the Einstein summation notation in (13). For ease of notation, the quantum stochastic terms for general function $\mathbf{V}(\rho_t)$ is denoted by $\aleph_{(\rho,t)}, \aleph_{(\rho,A)}, \aleph_{(\rho,A^\dagger)}, \aleph_{(\rho,\Lambda)}$. Then for a stop time $\tau$:

$$\mathbb{E}_\rho\left(\mathbf{V}(\rho_{t\wedge\tau})\right) = \mathbb{E}_\rho\left(\mathbf{V}(\rho)\right) + \int_0^{t\wedge\tau}\mathbb{E}_\rho\left(\aleph_{(\rho,t)}\right)dt + \int_0^{t\wedge\tau}\mathbb{E}_\rho\left(\aleph_{(\rho,A)}dA_t\right) + \int_0^{t\wedge\tau}\mathbb{E}_\rho\left(\aleph_{(\rho,A^\dagger)}dA_t^\dagger\right) + \int_0^{t\wedge\tau}\mathbb{E}_\rho\left(\aleph_{(\rho,\Lambda)}d\Lambda_t\right) \tag{14}$$

with a parallel argument to previous section, the vacuum expectation of quantum stochastic integrals is zero. By $\aleph_{(\rho,t)}$, there is several stability conditions for stochastic density matrix, which is implied in the following and since the proof is similar to previous section, hence we omit.

**Theorem 3.4 (stability of stochastic density operator)**: suppose there is positive function $\mathbf{V}(\rho_t)$ such that $\mathbb{E}_\rho(\mathbf{V}(\rho_e)) = 0, \mathbb{E}_\rho(\mathbf{V}(\rho_t)) > 0 : \forall \rho_t \in \mathbb{Q}_\varepsilon - \{\rho_e\}$ which $\mathbb{Q}_\varepsilon \triangleq \{\rho_t : \mathbf{V}(\rho_t) < \varepsilon \mathbf{I}\}$, then the equilibrium $\rho_e$ is:

1- Local quantum state stable if for initial density operator $\rho_0 \in \mathbb{Q}_\delta : \mathbb{E}_\rho(\aleph_{(\rho,t)}) \leq 0$, $\forall \rho_t \in \mathbb{Q}_\varepsilon$

2- Asymptotically quantum state stable if for initial density operator $\rho_0 \in \mathbb{Q}_\delta : \mathbb{E}_\rho(\aleph_{(\rho,t)}) < 0 : \forall \rho_t \in \mathbb{Q}_\varepsilon - \{\rho_e\}$

3- Exponentially quantum state stable if there is $a \in \mathbb{R}^+$ such that $\mathbb{E}_\rho(\aleph_{(\rho,t)}) + a\mathbb{E}_\rho(\mathbf{V}(\rho_t)) < 0 : \forall \rho_t \in \mathbb{Q}_\varepsilon - \{\rho_e\}$

## Acknowledgement

We appreciate Ramon Van Handel for some helpful discussion on stochastic process and to introducing some helpful references.